\documentclass[aps,prl,twocolumn,showkeys,superscriptaddress]{revtex4-1}

\usepackage{epsfig}
\usepackage{latexsym}
\usepackage{graphicx}
\usepackage{epsfig}
\usepackage{epstopdf}
\usepackage{latexsym}
\usepackage{subfigure}

\usepackage{amsmath}
\usepackage{color}

\usepackage{lineno}

\def\beq{\begin{equation}}
\def\eeq{\end{equation}}

\def\CL{{\rm CL}}
\def\CV{{\rm CV}}

\begin{document}

\title{Spatiotemporal correlation uncovers fractional scaling in cardiac tissue}

\author{Alessandro Loppini}
\affiliation{Department of Engineering, Campus Bio-Medico University of Rome, Via A. del Portillo 21, 00128, Rome, Italy}
\author{Alessio Gizzi$^*$} 
\affiliation{Department of Engineering, Campus Bio-Medico University of Rome, Via A. del Portillo 21, 00128, Rome, Italy}
\author{Christian Cherubini}
\affiliation{Department of Engineering, Campus Bio-Medico University of Rome, Via A. del Portillo 21, 00128, Rome, Italy}
\affiliation{International Center for Relativistic Astrophysics (I.C.R.A.), Campus Bio-Medico University of Rome, Via A. del Portillo 21, 00128 Rome, Italy}
\affiliation{ICRANet, Piazza delle Repubblica 10, I-65122 Pescara, Italy}
\author{Elizabeth M. Cherry}
\affiliation{School of Mathematical Sciences, Rochester Institute of Technology, 85 Lomb Memorial Drive, Rochester, NY (USA)}
\author{Flavio H. Fenton}
\affiliation{School of Physics, Georgia Institute of Technology, 837 State Street, Atlanta, Georgia (USA).}
\author{Simonetta Filippi}
\affiliation{Department of Engineering, Campus Bio-Medico University of Rome, Via A. del Portillo 21, 00128, Rome, Italy}
\affiliation{International Center for Relativistic Astrophysics (I.C.R.A.), Campus Bio-Medico University of Rome, Via A. del Portillo 21, 00128 Rome, Italy}
\affiliation{ICRANet, Piazza delle Repubblica 10, I-65122 Pescara, Italy}

\begin{abstract}
Complex spatiotemporal patterns of action potential duration have been shown to occur in many mammalian hearts due to a period-doubling bifurcation that develops with increasing frequency of stimulation. Here, through high-resolution optical mapping and numerical simulations, we quantify voltage length scales in canine ventricles via spatiotemporal correlation analysis as a function of stimulation frequency and during fibrillation. We show that i) length scales can vary from 40 to 20 cm during one to one responses, ii) a critical decay length for the onset of the period-doubling bifurcation is present and decreases to less than 3 cm before the transition to fibrillation occurs, iii) fibrillation is characterized by a decay length of about 1 cm. On this evidence, we provide a novel theoretical description of cardiac decay lengths introducing an experimental-based conduction velocity dispersion relation that fits the measured wavelengths with a fractional diffusion exponent of 1.5. We show that an accurate phenomenological mathematical model of the cardiac action potential, fine-tuned upon classical restitution protocols, can provide the correct decay lengths during periodic stimulations but that a domain size scaling via the fractional diffusion exponent of 1.5 is necessary to reproduce experimental fibrillation dynamics. Our study supports the need of generalized reaction-diffusion approaches in characterizing the multiscale features of action potential propagation in cardiac tissue. We propose such an approach as the underlying common basis of synchronization in excitable biological media. 
\end{abstract}


\maketitle

\begin{figure}[t]
\centering
\includegraphics[width=1\linewidth]{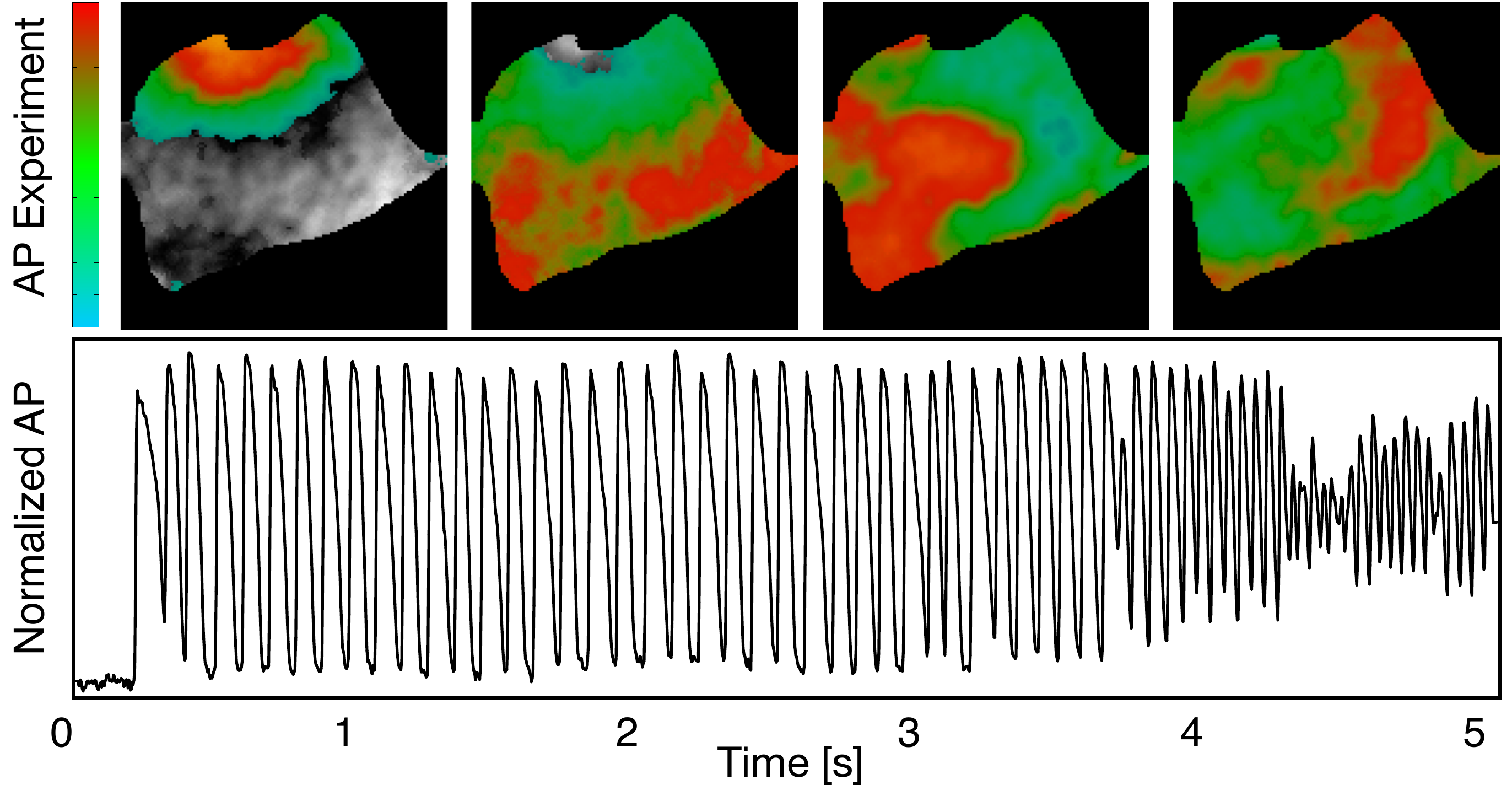}
\caption{Endocardial action potential (AP) voltage data from fluorescence optical mapping showing the transition from normal rhythm to ventricular tachycardia to ventricular fibrillation. Spatial distribution at selected frames (top) and time course of a single pixel (bottom).
Color code refers to normalized voltage amplitude. 
The grayscale background represents the endocardial ventricular tissue. See \cite{gizzi:2013} for details.}
\label{fig:fig1}
\end{figure}

The study of characteristic length scales in nonlinear dynamical systems, from astrophysics, fluid and plasma dynamics down to the cellular and molecular biology, has improved the predictive power of our mathematical descriptions.
The identification of large/small length scales as typical representations of long/short ranges of correlated dynamics within these systems has allowed us to identify the onset and to control critical behaviors, eventually avoiding their occurrence in physical scenarios~\cite{luther:2011}. In the realm of excitable biological media, the heart is a fascinating example of such complexity. It is intrinsically characterized by multiple spatial and temporal scales that trigger a broad spectrum of nonlinear dynamics and synchronization properties~\cite{glass:2001,qu:2014,scardigli:2017}. 
Cardiac arrhythmias, in particular, are well known to be supported by increased spatial dispersion of repolarization, inducing large variations in the refractory period and in the conduction velocity (CV)~\cite{han:1964,burton:2001}. These quantities have been linked to oscillations of the T-wave in the electrocardiogram signal, suggesting clinical importance in risk stratification for sudden cardiac death\,\cite{pastore:1999}.
The disorganization of ventricular electrical activity (fibrillation), in fact, represents an unpredictable condition arising from a complex series of spatiotemporal patterns, e.g., phase-locking and period-doubling bifurcations~\cite{guevara:1981} still requiring great effort to be fully understood. An experimental example of induction of fibrillation is provided in Fig.~\ref{fig:fig1} via fluorescence optical mapping of canine ventricular endocardium undergoing electrical stimulation~\cite{gizzi:2013}.

Attempts to classify these different regimes dates back to Wiggers~\cite{wiggers:1940} and this subject is widely studied in small animal experiments and isolated myocardium as well as supported by sophisticated mathematical models~\cite{cherry:2004,cherry:2008}. 
Although quantitative indicators have been proposed to identify the prevalent features during ventricular fibrillation~\cite{egolf:1994,bayly:1993}, generalized spatiotemporal descriptors characterizing the transition toward such a condition are still lacking. The concept of spatiotemporal chaos, in fact, in spatially extended systems was originally described by hyperbolic partial differential equations~\cite{cai:1999} and well-known applications regard the Kuramoto-Sivshinsky model~\cite{goren:1998} and the nonlinear Schr\"oedinger equations~\cite{cai:2001}, specifically addressing finite-size effects via decay mutual information statistics. Such dynamics require the appropriate control of the time and space organization of the system, e.g., fractal dimensions and Lyapunov exponents. In the context of cardiac dynamics, ventricular fibrillation has been shown to exhibit a strong periodic component centred near $8\,{\rm Hz}$~\cite{gray:1998}, seen as an attractor in two-dimensional-phase space, and indicating a consistent amount of temporal and spatial organization during cardiac fibrillation.

Here we propose a novel theoretical description of cardiac electrical properties based on the identification of characteristic lengths during paced excitation dynamics thus unveiling dispersive properties of the tissue. The methodology is verified both for experimental optical mapping recordings~\cite{gizzi:2013}, focusing on endocardial canine right ventricles, and for cardiac action potential simulations, making use of an accurate phenomenological mathematical model~\cite{fenton:2008,fenton:2013}. We base our study on the analysis of space-time correlation functions used for systems both at equilibrium and outside equilibrium~\cite{lumley:1970,panchev:1971,cross:1993} that in fluid mechanics allow the investigation of universal features within specific dynamical regimes, i.e., turbulence~\cite{frisch:1995}. We demonstrate that the characteristic length of the system reduces during fast pacing reaching the smallest value of about 1 cm during sustained fibrillation (no pacing). In this transition, we further identify the normalized decay length as a predictive indicator of period-doubling bifurcations (alternans). A major finding in this work, we provide a direct theoretical explanation for the observed behavior by introducing a phenomenological exponential law that replicates the characteristic length of the system and is in agreement with the fractional Laplacian operator showed to replicate experimental dispersion of repolarization in human cardiac tissue~\cite{bueno-orovio:2014,bueno-orovio:2015}. This new scaling law allows, in fact, to accurately reproduce and predict fibrillation behaviors by applying a domain size scaling process preventing from any additional model tuning (an extra step usually required to predict arrhythmias).

\begin{figure}[t]
\centering
\includegraphics[width=1\linewidth]{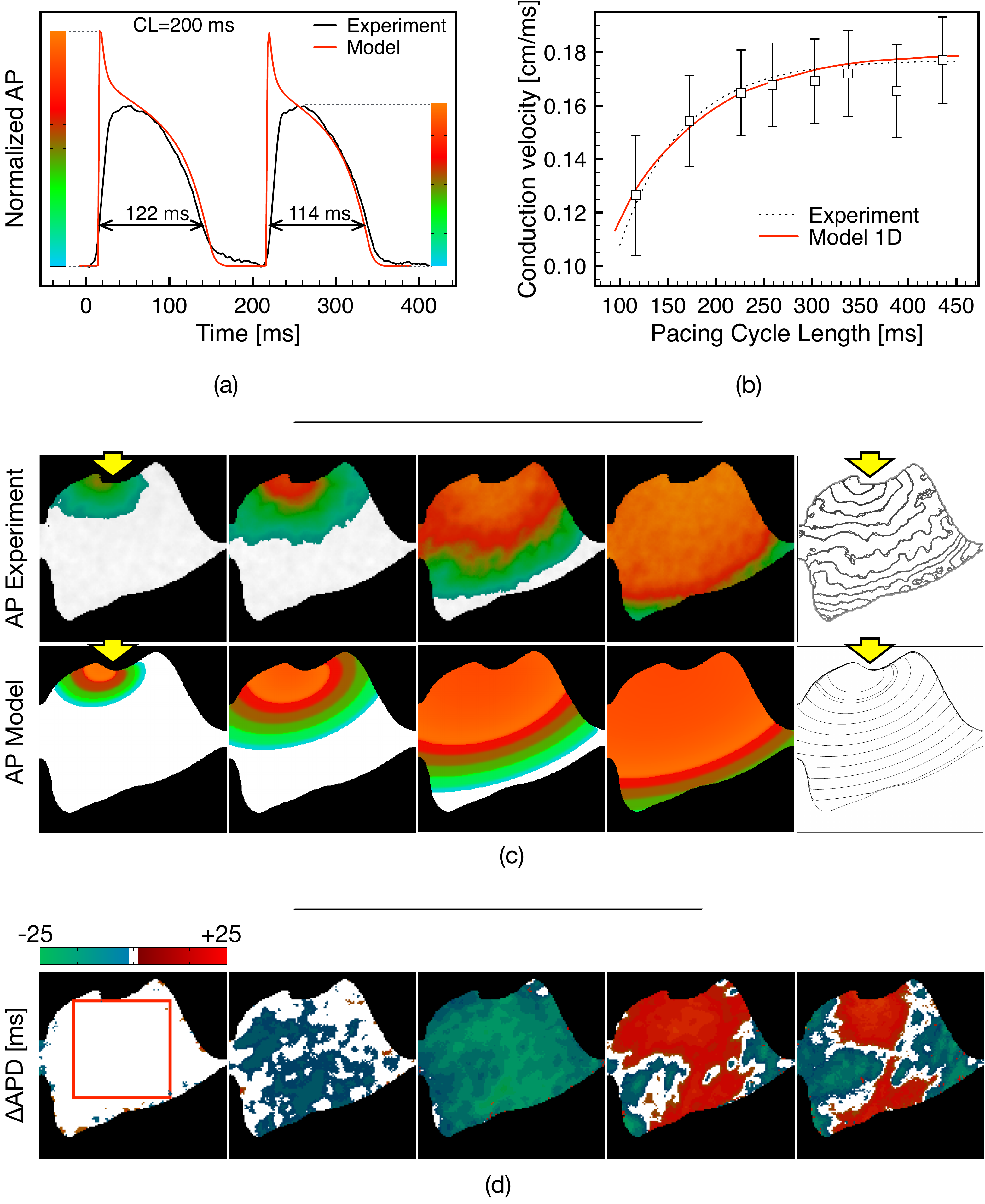}
\caption{\label{fig:fig2}
(a) Experimental (black) and simulated (red) action potential time course for two consecutive activations during fast pacing (CL=200 ms). Quantitative indication of APD alternans is provided.
(b) Conduction velocity restitution curves (mean plus standard error) for experimental tissue samples (symbol), fitting law (dashed), and one-dimensional model prediction (solid) as in Eq.~\ref{eq:CVCL}.
(c) Representative endocardial wavefront propagations and corresponding isochrones from experiments (top) and model (bottom). Yellow arrows indicate the location of the pacing electrode.
(d) Spatial map of $\Delta{\rm APD}$ alternans evolution during pacedown stimulation protocol~\cite{gizzi:2013} (see SI for details): non-alternating (white areas), concordant alternans (blue areas), discordant alternans (blue/red regions). Red square on the left indicates the region selected to compute the correlation function and to extract the decay length $L_0$ both for optical data and simulations.}
\end{figure}

\paragraph{Model Tuning.}
We show the complexity of the problem and the accuracy of the modeling approach by comparing temporal and spatial features classically adopted to fine-tune predictive computational models in cardiac electrophysiology~\cite{fenton:2008}. Figure~\ref{fig:fig2}(a) compares the time course of two consecutive action potentials during fast electrical pacing (cycle length--CL--equal to 200 ms) quantifying the action potential duration (APD) for a representative example of canine optical mapping recordings~\cite{gizzi:2013} and one-dimensional (1D) simulations. Figure~\ref{fig:fig2}(b) compares the conduction velocity (CV) calculated on the two-dimensional (2D) endocardial surface (average and standard error from 7 samples--squared symbols) for decreasing values of CL (restitution protocol~\cite{mironov:2008}) with respect to 1D model prediction (solid line). The 1D exponential fitting (dashed line) is also provided since, as clarified in the following section, it represents a fundamental feature of the system necessary for the \emph{a priori} estimate of the tissue dispersion during arrhythmic scenarios. Figure~\ref{fig:fig2}(c) shows experimental and simulated endocardial electrical excitations during a single action potential wave propagations clearly showing the accuracy of the numerical wave-front dynamics. In this case, the phase field approach is adopted~\cite{fenton:2005} such that the computational domain size corresponds to the irregular optical area taken from the measures. As an additional level of information, isochrones of activation are provided in both cases to highlight further the need of non-trivial anisotropies and heterogeneities in the computational model to reproduce the observed dynamics~\cite{barone:2017}. Finally, Fig.~\ref{fig:fig2}(d) shows a spatial view of alternans maps on the optical field of view (see Supplementary Material--SI) obtained on the same tissue for different pacing cycle lengths. In particular, from left to right, the CL decreases thus inducing a higher and more heterogeneous distribution of alternans in the tissue, up to 25 ms of APD difference for two consecutive beats. This analysis confirms the complex multiscale nature of the cardiac tissue and the intrinsic coupling between its spatial and temporal features.

\begin{figure}[t!]
\centering
	\subfigure{\includegraphics[width=0.48\textwidth]{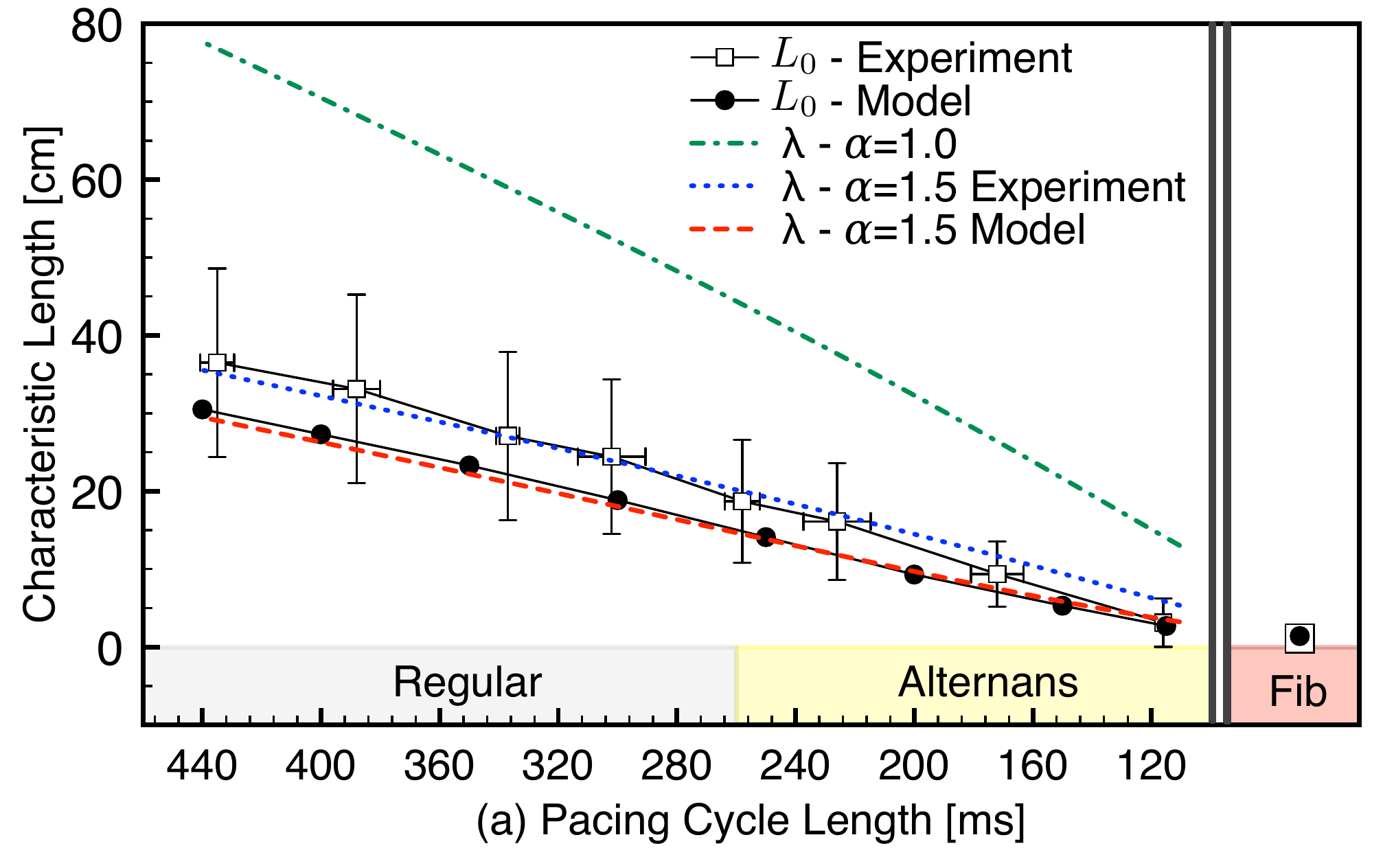}
	\label{fig:correlation}}
	\\
	\subfigure{\includegraphics[width=0.48\textwidth]{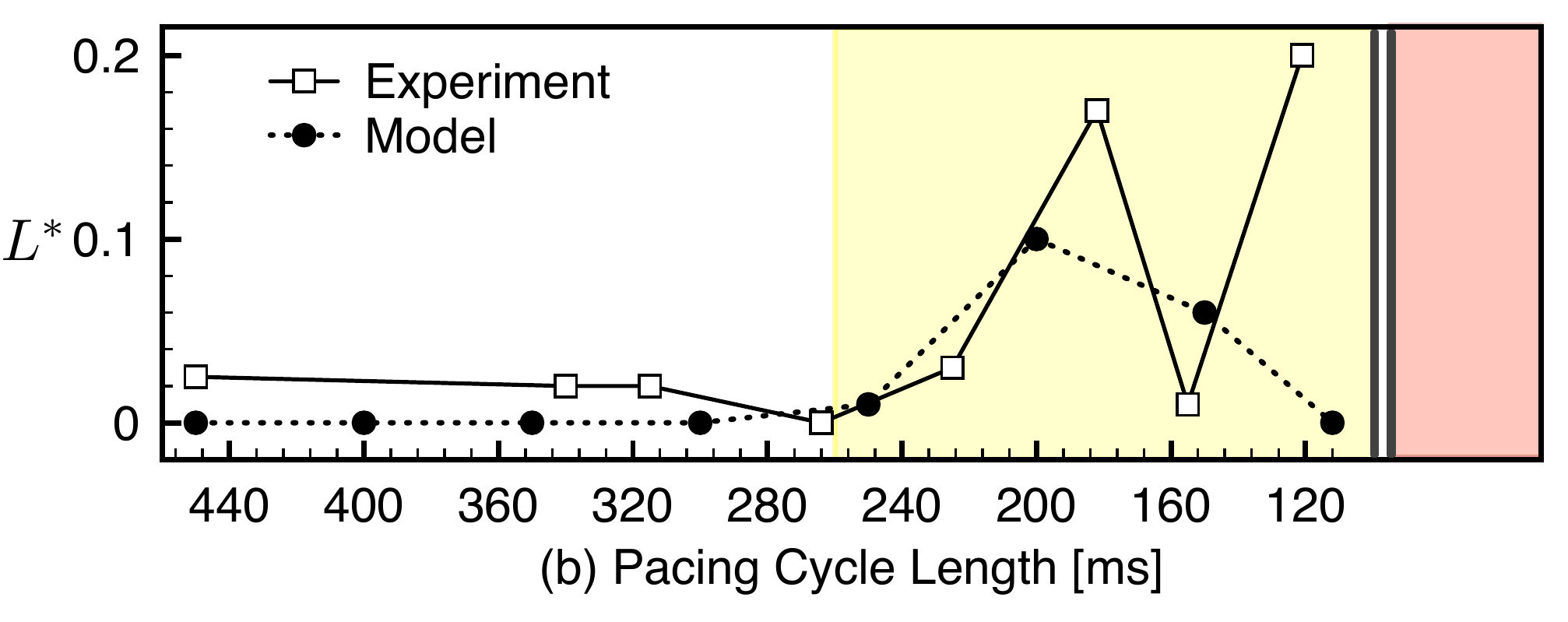}
\label{fig:correlationN}}
\caption{
(a) Decay length as a function of pacing cycle length for endocardial experimental recordings and corresponding model simulations. Standard deviation is superimposed to the experimental measures. Dashed curves indicate the fraction exponential relation, Eq.~\eqref{eq:CVCL}, in the idealized case ($\alpha=1$) and fitted for experiments and model ($\alpha=1.5$).
(b) Normalized decay length $L^*$ versus pacing cycle length for a representative experimental recording (solid) and model (dashed).
}
\end{figure}

\paragraph{Correlation Measure during Pacing.}
We introduce a novel quantitative analysis of fluorescence optical mapping signals based on the calculation of correlation functions and the identification of the corresponding characteristic spatial length (decay lengths--$L_0$--SI). Evaluated decay lengths are shown in Fig.~\ref{fig:correlation} for 7 different ventricular preparations. Optical data (squared symbols) are characterized by an average $L_0$ decreasing from 38 cm to 3 cm for the endocardial surface (34 to 4 cm for the epicardial surface, not shown), reducing the pacing cycle length from 450 to 115 ms. As expected, variability between different tissues is observed, but it decreases at short CLs where smaller $L_0$ values are identified, and significant exponential decay of the two-point correlation function is obtained. The robustness of the methodology is further confirmed by numerical simulations (filled circle symbols in Fig.~\ref{fig:correlation}) that match the experimental decay length trend finally merging the mean experimental value at short CLs.

To characterize the multiple transitions occurring at fast pacing, we enrich the previous analysis with the measure of the normalized decay length ($L^*$, see SI) as shown in Fig.~\ref{fig:correlationN}. This particular indicator can identify \emph{i)} the transition from no alternans to concordant alternans when a  net increase of $L^*$ is observed and \emph{ii)} the transition from concordant alternans to discordant alternans when consecutive oscillations of $L^*$ are present. First, for the experimental data at slow pacing rates (CL$>$300 ms), we observe that $L^*$ is not null thus implying an intrinsic dispersion in the tissue~\cite{gizzi:2013}. This aspect is of great importance to relate cardiac dynamics with theories of deterministic chaos in which the system is susceptible to initial conditions. Second, the complex discordant alternans case obtained for very short pacing rates (CL$<$150 ms) shows large oscillations of $L^*$ corresponding to the shortest $L_0$ in Fig.~\ref{fig:correlation}. Third, an intermediate \emph{re-synchronization} pattern is present and characterized by $L ^* \simeq0$. This particular condition is observed only as a critical state before a transition occurs. We theoretically justify these critical transitions in terms of $L^*$ values obtained via numerical simulation (dashed line in Fig.~\ref{fig:correlationN}). The model can reproduce the normalized decay length patterns both in amplitude and timing. In particular, null $L^*$ is obtained when alternans is not present in the tissue, and a net increase of $L^*$ appears for the same CL in which we observe alternans in the experiment. However, the model does not show any \emph{a priori} dispersion at slow pacing rates (i.e. constant $L^*\neq0$) and does not capture multiple oscillations at fast CLs. These two limitations are common in the current literature of mathematical physiology that should be enriched with memory effects and a priori dispersion, and that usually requires larger simulation domain to reproduces realistic alternans patterns.

\paragraph{Model-Based Dispersive Theory.}
By the provided model accuracy, we assume that the wave velocity varies with the pacing period according to the exponential law $\CV(\CL)=a-b\exp{(c \, \CL)}$. Accordingly, we fit the experiments with parameters $a=0.177,b=0.31,c=-0.015$ (see Fig.~\ref{fig:fig1}(b) dashed line). Also, we define the characteristic length of the excitation wave, $\lambda$, introducing a new phenomenological dispersion relation in which the pacing CL acts like an internal variable:
\beq\label{eq:CVCL}
	\lambda 
	= 
	\left[\dfrac{\CV(\CL)}{{\rm CV}_0}\right]^{\alpha} ( {\rm CV}_0  \cdot \CL )
	\equiv
	\left[a-b \,{\rm e}^{(c \, \CL)}\right]^{\alpha}  \CL.
\eeq
Here, $a,b$ [${\rm (cm/ms)}^{1/\alpha}$] and $c$ [$1/{\rm ms}$] describe the experimental/model-based CV(CL) relationships. Such a relation has the notable limit of linear wave propagation for $\alpha=1$, which represents the steady-state value of the wavelength when the system is non-dispersive (we verified this limit via 1D cable simulations of length 100 cm). The result is shown in Fig.~\ref{fig:correlation} (green dashed line) using the $\CV(\CL)$ experimental fit curve posing $\alpha=1$. However, a notable difference appears for the values of $\lambda$ obtained from experiments (dashed blue) and model (dashed red) $\CV(\CL)$ relationships. Our analysis shows that both for the experiment and model laws the dispersive exponent able to replicate the measured decay lengths corresponds to $\alpha=1.5$. This particular value is in agreement with the fractional Laplacian operator exponent showed to best replicate experimental dispersion of repolarization in human cardiac tissue~\cite{bueno-orovio:2014,bueno-orovio:2015} and that is based on a microscopic biophysics description of cardiac propagation. The result, by analogy, can be read as \emph{i)} a spatiotemporal generalization of scale invariance usually adopted in fractal geometry~\cite{goldberger:2002} and \emph{ii)} the fractional diffusion description of cell-cell coupling in cardiac electrophysiology~\cite{bueno-orovio:2014,bueno-orovio:2014a,bueno-orovio:2016}.

\paragraph{Analysis and Simulation of Ventricular Fibrillation.}
Ventricular fibrillation is finally analyzed both for the experimental preparations and the mathematical model. Such an auto-excitatory regime presents multiple unstable spirals at the same time within the tissue (usually three for both experiments and simulations--see SI). In this scenario, the external excitation pacing is not present anymore, and much shorter decay lengths characterize the system. Our analysis reveals that $L_0$ falls to an average value of 1.1 cm for the endocardial experimental data (see Fig.~\ref{fig:correlation} 
red area--squared symbol). Accordingly, numerical simulations confirm the short $L_0$ value of about 1.4 cm when a domain scaling procedure is applied (see Fig.~\ref{fig:correlation} red area--circle symbol). Specifically, since the mathematical model is tuned upon CV restitution properties (Fig.~\ref{fig:fig2}(b)), we introduced the following procedure: (1) we simulate a sustained fibrillation scenario in a large squared domain with \emph{fictitious} side length of $\Delta=20$ cm; (2) we perform a down-scaling of the domain size according to the CV(CL) dispersive law obtaining a \emph{physical} domain size of $\delta=\Delta^{1/\alpha}=7.4$ cm by using $\alpha=1.5$; (3) we apply the spatiotemporal correlation analysis within the \emph{physical} box of size $3\times3$ cm$^2$ (in agreement to the experimental case) and identify the sought decay length. Decay lengths of the order of $\sim1$ cm provide evidence that cardiac fibrillation is not a spatially random mechanism but a high-dimensional process characterized by a measurable degree of coherence\,\cite{goldberger:1986,bayly:1993,gray:1998,yashima:2003}.

\paragraph{Conclusions and Perspectives.}
Ventricular synchronization features are associated with a long-range correlated dynamics over the whole tissue size corresponding to long spatial wavelengths of depolarization/repolarization. Small values, instead, are usually associated with coherent local states. However, the interpretation of spatial correlation is not restricted to a certain physical system nor linked to a particular mechanism (i.e., spiral waves)~\cite{bayly:1993}. In the context of cardiac dynamics, previous studies proposed to apply correlation functions as a tool to analyze cardiac fibrillation~\cite{kaplan:1990,bayly:1998}. The chaotic nature of electrochemical waves in the heart and the quantification of the degree of the spatial organization during ventricular fibrillation, in fact,  are fundamental for understanding and developing useful models finding effective clinical treatments. 

In this work we adopted similar reasoning to study spatial scales of correlations characterizing the transition of the excitation wave starting from normal rhythm (non-alternating), passing through a period-doubling bifurcation (concordant and discordant alternans) and ending with sustained ventricular fibrillation. We found that a long-range interaction (high values of $L_0$) is characteristic of longer CLs. The physical meaning of such a value is regarded as the total length through which the activation wave must propagate to correctly synchronize the whole organ as well as to restore the resting condition in a unified manner (full depolarization and repolarization phases). This result is further confirmed by the fact that endocardial and epicardial surfaces do not present significant differences in $L_0$. On these bases, following the usual restitution procedure in cardiac electrophysiology, we quantified the correlation length transitions obtained during pacedown stimulation protocol recorded for several tissue preparations~\cite{gizzi:2013}. In particular, we were able to identify critical values of the decay length: i) at the onset of discordant alternans (for CL $\sim200$ ms) we detected $L_0\sim10$ cm; ii) at the onset of fibrillation (for CL $\sim100$ ms) we found $L_0<3$ cm.

The analysis of optically mapped canine ventricular preparations was theoretically confirmed via fine-tuned two-dimensional numerical simulations. In this regard, it is worth mentioning that the usual modeling approaches base the parameters' identification according to local dynamics of the action potentials, in particular, APD restitution properties. However, due to the nonlinearity of the reaction-diffusion system, multiple combinations of the model parameters reproduce the same APD restitution properties but indeed modify the CV characters emerging from the spatiotemporal problem. Based on our theory, this, in turn, affects the resulting correlation dynamics and the characteristic length scale of the system reducing the predictive power of the model. Therefore, as the major aim of the present work, we introduced a new dispersive phenomenological law relating conduction velocity, wavelength and pacing period. As such, the dispersive relation is characterized by a fractional exponent ($\alpha=1.5$) found to replicate best the experimental decay length values obtained via the spatiotemporal correlation analysis. The selected fractional exponent is then extrapolated to fibrillation scenarios introducing a novel scaling law of the simulated domain size. This new procedure allowed us to reproduce the physical spatiotemporal features of the system without modifying any the model parameter. Such theoretical reasoning, making use of a nonlinear dispersive law, clearly suggests the need of a deeper understanding of cardiac tissue microstructural features and non-local diffusion operators~\cite{hurtado:2016,cherubini:2017,cusimano:2018} to replicate emerging phenomena in cardiac electrophysiology with a significant interest in unveiling micro-scale defects leading to pathological states.

Authors acknowledge the support of the International Center for Relativistic Astrophysics Network (ICRANet), the Italian National Group for Mathematical Physics (GNFM-INdAM), the Visiting Professor Programme at Campus Bio-Medico University of Rome, the National Science Foundation under grant no. CNS-1446312.

\bibliographystyle{apsrev4-1} 
%

\section{Supplementary Material}
\subsubsection{\bf Mathematical Model.}

We consider the four-variable Minimal Model of ventricular action potential~\cite{fenton:2008,fenton:2013} tuned to match optical-mapping data~\cite{gizzi:2013} (see Fig.~\ref{fig:fig5}) recovering restitution properties and activation patterns:

\begin{subequations}
\begin{align}
\label{eq:MM_MODEL1} 
\partial_t u &= \nabla \cdot \left( D_{ij}\nabla u \right) -
\left(J_{fi}+J_{so}+J_{si}\right)
\\
\label{eq:MM_MODEL2} 
\partial_t v &=
\left[1-H(u-\theta_v)\right]\frac{(v_\infty-v)}{\tau_v^-} -
\frac{H(u-\theta_v)v}{\tau_v^+}
\\
\label{eq:MM_MODEL3} 
\partial_t w &=
\left[1-H(u-\theta_w)\right]\frac{(w_\infty-w)}{\tau_w^-} -
\frac{H(u-\theta_w)w}{\tau_w^+}
\\
\label{eq:MM_MODEL4} 
\partial_t s &=
\frac{(1+\tanh(k_s(u-u_s)))/2 - s}{\tau_s}\,,
\end{align}
\end{subequations} 
where $u$ is the dimensionless membrane potential ($u$ can be rescaled to dimensions of $\rm mV$ by using the map $V_m=(85.7u-84)\,{\rm mV}$), and $v$, $w$, $s$ are the gating variables modeling activation and inactivation of ions currents; $H(x)$ represents the classical Heaviside step function; $J_{fi}$, $J_{so}$, $J_{si}$ denote fast-inward sodium, slow-outward potassium and slow-inward calcium density currents, whose expressions are:

\begin{subequations}
\begin{align}
\label{eq:MM_MODEL5} J_{fi} &=
-H(u-\theta_v)(u-\theta_v)(u_u-u)\frac{v}{\tau_{fi}}
\\
\label{eq:MM_MODEL6} J_{so} &=
\left[1-H(u-\theta_w)\right]\frac{(u-u_o)}{\tau_o}+\frac{H(u-\theta_w)}{\tau_{so}}
\\
\label{eq:MM_MODEL7} J_{si} &=
-H(u-\theta_w)\frac{ws}{\tau_{si}}\,.
\end{align}
\end{subequations}

The voltage-dependent time constants and asymptotic values are given by:

\begin{subequations}
\begin{align}
\label{eq:tau1}
\tau_v^- (u) &= \left[1-H(u-\theta_v^-)\right]\tau_{v1}^- + H(u-\theta_v^-)\tau_{v2}^-\\
\label{eq:tau2}
\tau_w^+ (u) &= \tau_{w1}^+ + (\tau_{w2}^+ - \tau_{w1}^+)\frac{\tanh(k_w^+(u-u_w^+))+1}{2}\\
\label{eq:tau3}
\tau_w^- (u) &= \tau_{w1}^- + (\tau_{w2}^- - \tau_{w1}^-)\frac{\tanh(k_w^-(u-u_w^-))+1}{2}\\
\label{eq:tau4}
\tau_{so} (u) &= \tau_{so1} + (\tau_{so2} - \tau_{so1})\frac{\tanh(k_{so}(u-u_{so}))+1}{2}\\
\label{eq:tau5}
\tau_s (u) &= \left[1-H(u-\theta_w)\right]\tau_{s1} + H(u-\theta_w)\tau_{s2}\\
\label{eq:tau6}
\tau_o (u) &= \left[1-H(u-\theta_o)\right]\tau_{o1} + H(u-\theta_o)\tau_{o2}\\
\label{eq:inf}
v_{\infty} &= \left\{ \begin{array}{rr}
1, & u<\theta_v^-\\ 
0, & u\ge\theta_v^-
\end{array}\right.\\
w_{\infty} &= \left[1-H(u-\theta_o)\right]\left(1-\frac{u}{\tau_{w\infty}}\right) + H(u-\theta_o)w_{\infty}^*\,.
\end{align}
\end{subequations}

The two-dimensional diffusion tensor $D_{ij}$ is defined concerning the conductivity tensor $\sigma_{ij}$, the cell surface to volume ratio $S_o$ and the membrane capacitance  $C_m$:

\begin{equation}
{D}_{ij} \equiv \frac{\sigma_{ij}}{S_o\,C_m} = \left(
\begin{array}{cc}
D_{11} & D_{12}\\
D_{21} & D_{22}\\
\end{array}
\right)\,,
\end{equation}
where diagonal and off-diagonal elements are defined based on diffusivities along directions parallel and orthogonal to the fibers, and considering fibers angle rotation $\theta\left(x,y\right)$:

\begin{subequations}
\begin{align}
D_{11} &= D_{\parallel} \cos^2\left(\theta\left(x,y\right)\right) + D_{\bot} \sin^2\left(\theta\left(x,y\right)\right) \,,\\
D_{22} &= D_{\parallel} \sin^2\left(\theta\left(x,y\right)\right) + D_{\bot} \cos^2\left(\theta\left(x,y\right)\right) \,,\\
D_{12} &= D_{21} = \left(D_{\parallel}-D_{\bot}\right)\cos\left(\theta\left(x,y\right)\right) \sin\left(\theta\left(x,y\right)\right) \,.
\end{align}
\end{subequations}

\begin{figure}[h!]
\centering
	{\includegraphics[width=0.45\textwidth]{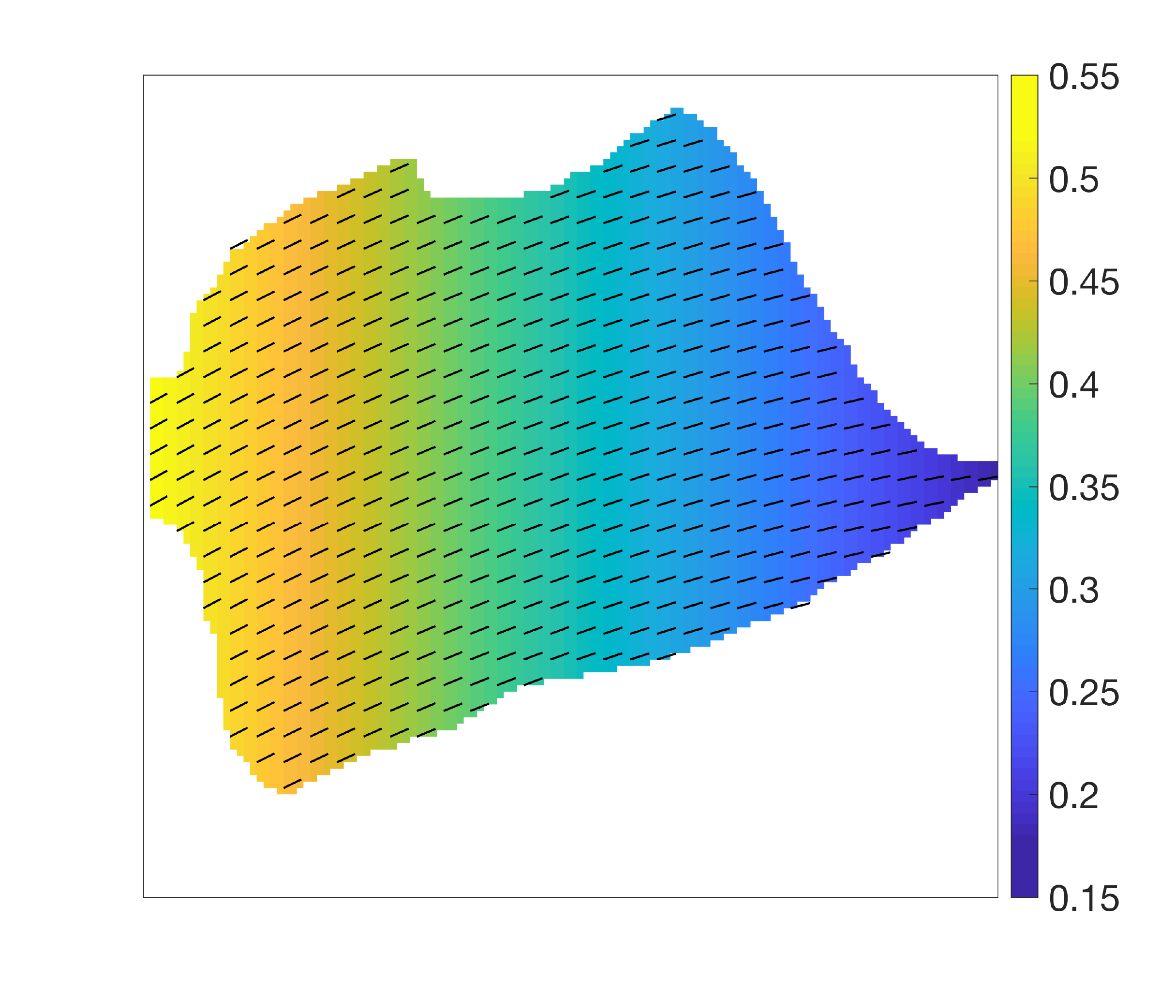}}
\caption{Representative example of spatial distribution of the fiber angular variation according to Eq.~\eqref{eq:fibers}.
\label{fig:fig00}
} 
\end{figure}

\begin{figure*}[h!]
\centering
	\subfigure{\includegraphics[width=0.95\textwidth]{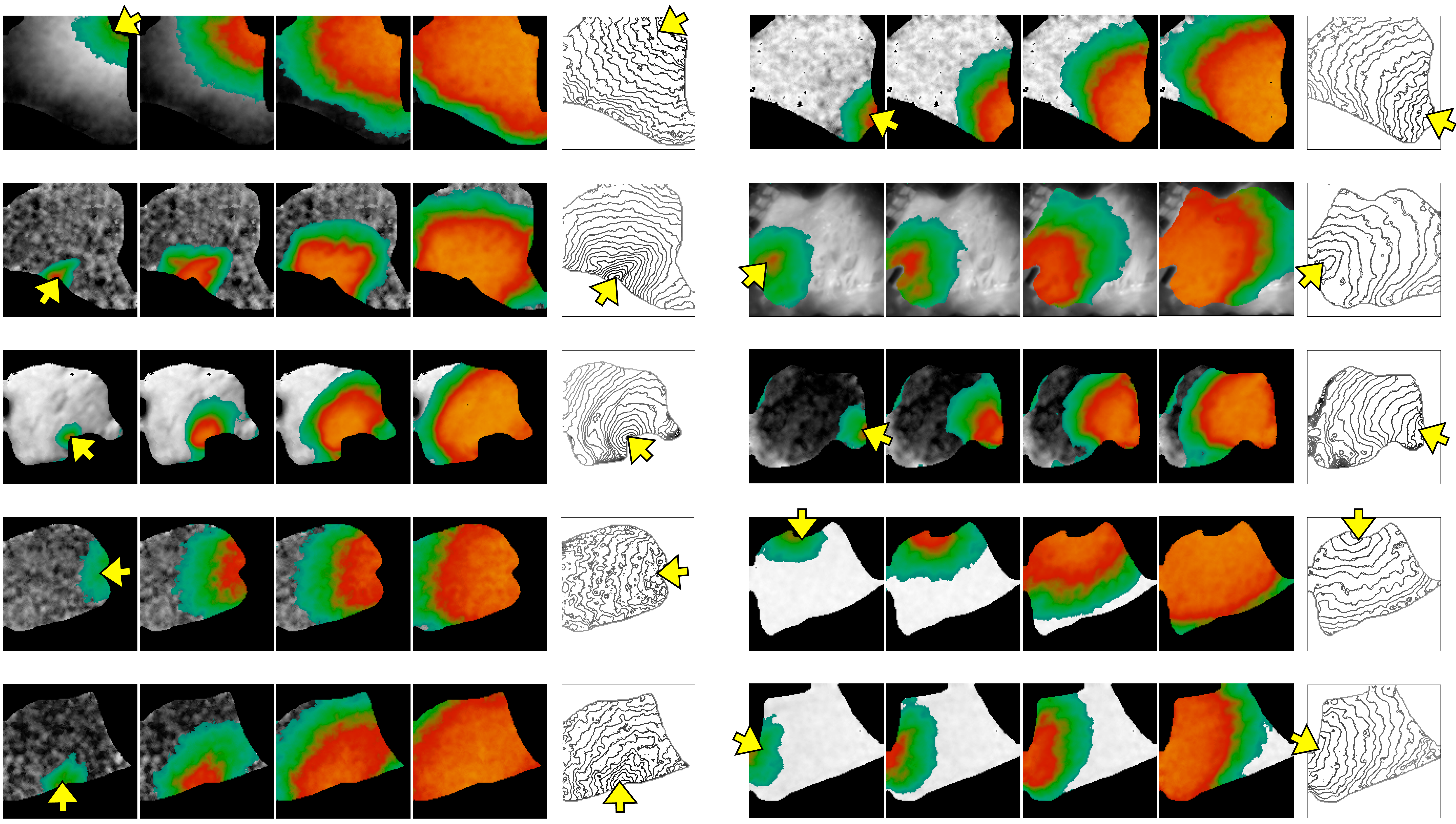}}
\caption{Collection of ventricular wedge electrical stimulation under different pacing sites~\cite{gizzi:2013}. Activation isochrones are provided on the right for each specimen.
\label{fig:fig5}
} 
\end{figure*}

The angular variation $\theta(x,y)$ is fine-tuned for each specimen shown in Fig.~\ref{fig:fig00}, by using the following linear functions:

\begin{subequations}\label{eq:fibers}
\begin{align}
\theta(x,y)&=\theta(0,y) + (\theta(L,y)-\theta(0,y))\,x/L \,,\\
\theta(x,y)&=\theta(x,0) + (\theta(x,L)-\theta(x,0))\,y/L \,,
\end{align}
\end{subequations}
where $x=j\Delta x$, $y=i\Delta y$ and $L=n_x\Delta x=n_y\Delta y$, assuming a mesh size $\Delta x=\Delta y$ and $n_x=n_y$ mesh points.

Model parameters are provided in Tab.1.

Numerical simulations were solved on two-dimensional domains via a finite-difference discretization of spatial derivatives and explicit time-integration via the forward Euler method. We adopted $dx=0.025$ cm and $dt=0.01$ ms. Such choice ensures sufficiently accurate results, checked against finer meshes, and is in line with previous computational works (see e.g. Fenton \& Karma 1998). Conduction velocity values were further taken into account when analyzing numerical convergence.

\subsubsection{\bf APD Alternans.}
The spatial distribution of the alternating phases of the action potential across the mapped field was defined as the difference between two consecutive action potential duration (APD):

\begin{eqnarray}
\label{eq:alternans}
\Delta {\rm APD}(x,y)_n
=
{\rm APD}(x,y)_{n+1} - {\rm APD}(x,y)_n
\rightarrow \nonumber
\\
\left\{
  \begin{array}{l l}
    |\Delta{\rm APD}(x,y)_n| > 2\,ms & \quad \text{Alternans}\\
    |\Delta{\rm APD}(x,y)_n| \leq 2\,ms & \quad \text{Nodal line}\\
    \end{array} \right.
\end{eqnarray}
where $n$ denotes the beat number and ${\rm APD}(x,y)$ is the duration of the action potential at a pixel in position $(x,y)$ within the 2D mapped field. Tissue is defined non-alternating if $|\Delta{\rm APD}(x,y)_n|<2$ms 
and alternating otherwise.
We refer to~\cite{gizzi:2013} for further details of the experimental setup.

\subsubsection{\bf Correlation Analysis.}
\begin{figure*}[h!]
\centering
	\subfigure[Experiment]{\includegraphics[width=0.3\textwidth]{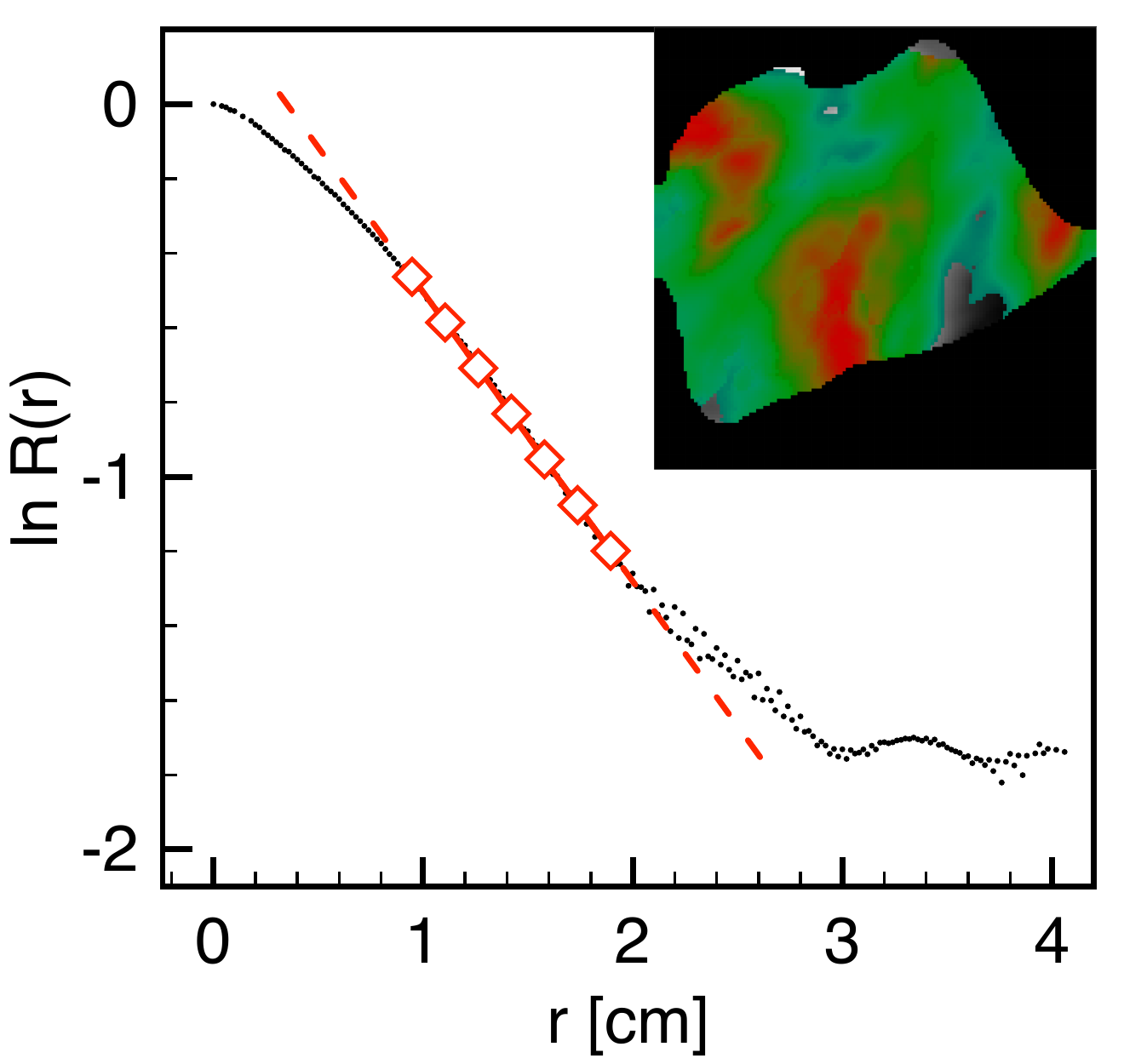}}
	\subfigure[Model]{\includegraphics[width=0.3\textwidth]{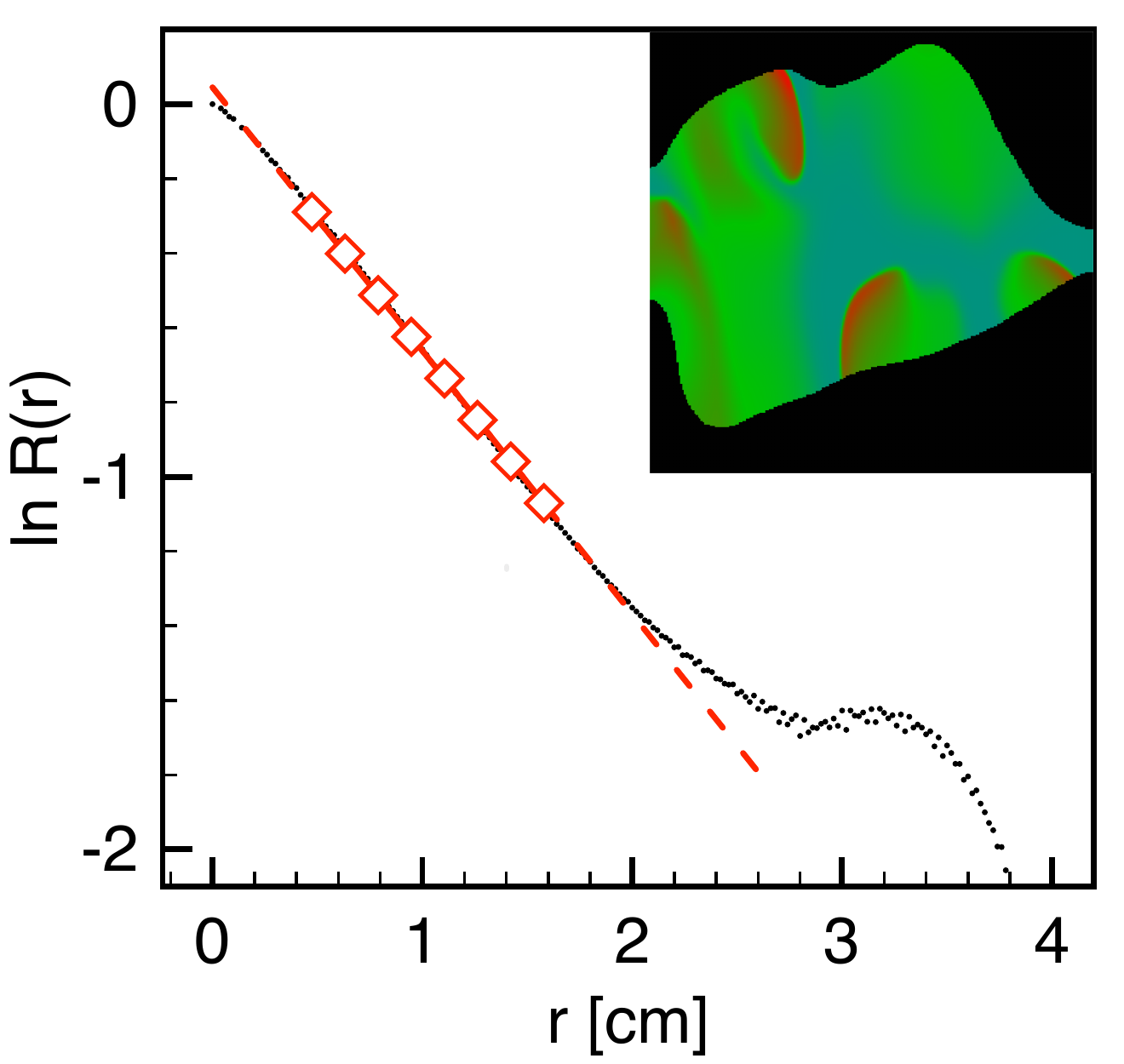}}
\caption{
Two-point correlation function during experimental (a) and model (b) fibrillation regimes. The insets provide a representative frame with multiple spirals within the region of interest. 
Red line represents the selected slope and symbols the decay length evaluation region: $L_0\simeq1.3\,cm$ (experiment), $L_0\simeq1.5\,cm$ (model), respectively.
\label{fig:fig4}
} 
\end{figure*}

Our analysis consists in evaluating the spatiotemporal decay length~\cite{egolf:1994,bayly:1993,bayly:1998,cai:2001,frisch:1995} during electric stimulation for decreasing pacing periods (pacedown protocol) and during sustained fibrillation. A two-dimensional mapped field is selected within a square region of about $3\times3$ cm$^2$ for both optical data and numerical simulations. 
A two-point correlation function is computed as the product-moment correlation index $R(\vec{r})$ between voltage time series extracted from selected points at distance $\vec{r}$:

\begin{equation}
\label{eq:corrL}
   R(\vec{r}) = 
   \frac{{\rm cov}(V_A,V_B)}{\sigma_A \sigma_B} = 
   \frac{\langle (V_A-\langle V_A \rangle) (V_B-\langle V_B \rangle) \rangle}{\sigma_A \sigma_B}\,.
\end{equation}
Here, $V_A=V(\vec{x},t)$, $V_B=V(\vec{x}+\vec{r},t)$ and $\langle\cdot\rangle$ denotes the time average operator computed within the selected time window and $\sigma_A,\sigma_B$ are the standard deviations of $V_A$ and $V_B$ processes. In this analysis, we considered two consecutive activation potentials for each pacing period. The correlation index is finally averaged over the spatial region of interest. Information about the decay length is extracted from the linear range in the semi-log plot of the correlation index (see Fig.~\ref{fig:fig4}) at increasing distances between selected points, i.e.:

\begin{equation}
   R(\vec{r})\propto \exp\left(-{\|\vec{r}\|}/{L_0}\right),
\end{equation}
where $L_0$ represents the sought decay length.

A second two-point correlation measure
(not shown) was also computed corroborating the results given by \eqref{eq:corrL}. For the two consecutive APs (+, -) we computed the normalized decay length $L^*$ based on the average value $\langle L_0 \rangle$ defined on the single decay length ($L_0^+, L_0^-$), i.e.:

\beq
\label{eq:DLN}
	\langle L_0 \rangle = \dfrac{L_0^+ + L_0^-}{2},
	\quad
	L^*= \dfrac{L_0^+ - \langle L_0 \rangle}{\langle L_0 \rangle},
\eeq
which is able to identify period-doubling regimes occurring in the system (a symmetric behavior is obtained by using $L_0^-$ in Eq.~\eqref{eq:DLN}, by definition).

\begin{table*}[h]
\label{tab:params}
Tab.1. Model parameters for endocardial action potential MM formulation.
Units are given in $\rm ms$, $\rm cm$, $\rm mV$, $\rm cm^2/ms$.\\ 
\begin{tabular}{lllll}
\\
\toprule
{\scriptsize $u_o =0$} &
{\scriptsize $\tau_{v2}^{-}=40$} &
{\scriptsize $u_{w}^{+}=0.0005$} &
{\scriptsize $\tau_{s1}=2.7342$} &
{\scriptsize $D_{\parallel}=0.010$}
\\
{\scriptsize $u_u =1.56$} &
{\scriptsize $\tau_{w1}^{-}=40$} &
{\scriptsize $\tau_{fi}=0.10$} &
{\scriptsize $\tau_{s2}=2$} &
{\scriptsize $D_{\bot}=0.003$}
\\
{\scriptsize $\theta_{v} =0.3$} &
{\scriptsize $\tau_{w2}^{-}=115$} &
{\scriptsize $\tau_{o1}=470$} &
{\scriptsize $k_{s}=2.0994$}
\\
{\scriptsize $\theta_{w} =0.13$} &
{\scriptsize $\tau_{w1}^{+}=175$} &
{\scriptsize $\tau_{o2}=6$} &
{\scriptsize $u_{s}=0.9087$}
\\
{\scriptsize $\theta_{v}^{-} =0.2$} &
{\scriptsize $\tau_{w2}^{+}=230$} &
{\scriptsize $\tau_{so1}=40$} &
{\scriptsize $\tau_{si}=2.9013$}
\\
{\scriptsize $\theta_{o} =0.006$} &
{\scriptsize $k_{w}^{-}=20$} &
{\scriptsize $\tau_{so2}=1.2$} &
{\scriptsize $\tau_{w \infty}=0.0273$}
\\
{\scriptsize $\tau_{v}^{+}=1.4506$} &
{\scriptsize $u_{w}^{-}=0.00615$} &
{\scriptsize $k_{so}=2$} &
{\scriptsize $w_{\infty}^{*}=0.78$}
\\
{\scriptsize $\tau_{v1}^{-}=55$} &
{\scriptsize $k_{w}^{+}=8$} &
{\scriptsize $u_{so}=0.65$} &
\\
\toprule
\end{tabular}
\end{table*}

\end{document}